\documentclass{iaus}
\usepackage{graphicx}

\usepackage{natbib}

\def\apj{ApJ}

\def\apjl{ApJ}
\def\mnras{MNRAS}
\def\araa{ARA\&A}

\newcommand{\mbh}{\ensuremath{M_\mathrm{BH}}}

\newcommand{\msun}{\,{\rm M_\odot}}

\newcommand{\msigma}{M_{\rm BH}-\sigma}

\newcommand{\beq}{
\begin{equation}
}
\newcommand{\eeq}{
\end{equation}
}
\newcommand{\ba}{
\begin{eqnarray}
}
\newcommand{\ea}{
\end{eqnarray}
}
\def\spose#1{\hbox to 0pt{#1\hss}}
\newcommand{\lta}{\mathrel{\spose{\lower 3pt\hbox{$\mathchar"218$}}
      \raise 2.0pt\hbox{$\mathchar"13C$}}}
\newcommand{\gta}{\mathrel{\spose{\lower 3pt\hbox{$\mathchar"218$}}
      \raise 2.0pt\hbox{$\mathchar"13E$}}}
\newcommand{\simlt}{{\mathrel{\rlap{\lower 3pt\hbox{$\sim$}}\raise 2.0pt\hbox{$<$}}}}
\newcommand{\simgt}{{\mathrel{\rlap{\lower 3pt\hbox{$\sim$}} \raise 2.0pt\hbox{$>$}}}}

\title[IAUS 267.~~Evolution of black holes] 
{The early evolution of massive black holes}

\author[Marta Volonteri]   
{Marta Volonteri$^1$}

\affiliation{$^1$ Astronomy Department, University of Michigan\\ 500 Church Street, Ann Arbor, MI 48109, USA\\ email: {\tt martav@umich.edu}}

\pubyear{2009}
\volume{267}  
\pagerange{119--126}
\jname{Co-Evolution of Central Black Holes and Galaxies}
\editors{B.M.\ Peterson, R.S.\ Somerville, \& T.\ Storchi-Bergmann, eds.}
\begin{document}

\maketitle

\begin{abstract}
Massive black holes are nowadays believed to reside in most local galaxies.   Studies have also established a number of relations between the MBH mass and properties of the host galaxy such as bulge mass and velocity dispersion. These results suggest that central MBHs, while much less massive than the host ($\sim$ 0.1\%), are linked to the evolution of galactic structure. When did it all start? In hierarchical cosmologies, a single big galaxy today can be traced back to the stage when it was split up in hundreds of smaller components. Did MBH seeds form with the same efficiency in small proto-galaxies, or did their formation had to await the buildup of substantial galaxies with deeper potential wells? I briefly review here some of the physical processes that are conducive to the evolution of the massive black hole population. I will discuss black hole formation processes for `seed' black holes that are likely to place at early cosmic epochs, and possible observational tests of these scenarios. 

\end{abstract}

\firstsection 
\section{Introduction}

I will focus here on the formation and evolution of massive black holes (MBHs), in high-redshift galaxies, and their symbiotic evolution with their hosts.  MBHs weighing million solar masses and above have been recognized as the engines that power quasars detected at early cosmic times.  Dynamical evidence also indicates that MBHs with masses in the range $M_{BH} \sim 10^6-10^9\,\msun$ ordinarily dwell in the centers of most nearby galaxies \citep{ferrareseford}. MBHs populate galaxy centers today, and shone as quasars in the past; the quiescent MBHs that we detect now in nearby bulges would be the dormant remnants of this fiery past, suggesting a single mechanism for assembling MBHs and forming galaxies.  The surprisingly clear correlations between MBH masses and the properties of their host galaxies \citep[and references therein]{Gultekin2009} may well extend down to the smallest masses, for example, the dwarf Seyfert~1 galaxies POX  52 and NGC 4395 are thought to contain a MBH of mass $M_{BH} \sim 10^5\,M_\odot$ \citep{barthetal2004,Peterson2005}.  At the other end, however, some powerful quasars, powered by super-massive black holes  with masses $\simeq 10^9\, M_\odot$ \citep{Barthetal2003}, have already been detected at $z>6$, corresponding to a time less than a tenth of the age of the Universe, roughly one billion years after the Big Bang. 

We are therefore left with the task of explaining the presence of  billion solar masses MBHs when the Universe is less than {\rm 1 Gyr} old,  and of much smaller MBHs lurking in {\rm 13 Gyr} old galaxies. The outstanding questions concern  how and when ``seed'' MBHs formed, the frequency of MBHs in galaxies, and how efficiently MBH seeds grew in mass during the first few billion years of their lives.

The ``flow chart" presented by \cite{Rees1978} still stands as a guideline for the possible paths leading to formation of massive MBH seeds in the center of galactic structures.  In this paper I will concentrate on the currently favoured scenarios.


%

\section{Massive black hole `seeds'}

\subsection{Population III remnants}
\label{ssec:1a}

One of the most popular scenarios for MBH formation associates MBH seeds with the remnants of the first generation of stars, formed out of zero metallicity gas.  The first stars are expected to form in minihalos, $M_{\rm min}\approx  10^6\,\msun$ collapsing at $z\sim 20-50$ from the highest peaks of the primordial density field, where cooling is possible by means of molecular hydrogen \citep{palla2002}.   Simulations of the collapse of primordial  molecular clouds \citep{bromm1999,bromm2002,abel2000,Yoshida2006}  suggest that the first generation of stars contained many `very massive stars' (VMSs) with $m_\star>100\,\msun$.  This is because of the slow subsonic contraction set up by molecular hydrogen cooling. Further fragmentation into sub-components is not seen although it is not clear if this is  a numerical effect,  rather than due to the gas physics \citep{Glover2008a}. 

If the first stars retain their high mass until death, they will collapse after a short ($\approx$ Myrs) life-time. The final fate depends on the exact mass of the star.  Between 25 and 140 $\msun$, low-metallicity stars form black holes directly. It is likely, however, that such light black holes wander within their hosts, dynamically interacting with stars of similar mass, rather than settle at the center of the galaxy's potential well.  Between approximately 140 and 260 $\msun$ lies the domain of pair instability supernovae. These objects are completely disrupted by nuclear-powered explosions, leaving no remnants  \citep{Kudritzki2000}. Going to still more massive stars (over 260 $\msun$) the nuclear energy released when the star collapses on the pair instability is not sufficient to reverse the implosion. A massive black hole, containing at least half of the initial stellar mass, is born inside the star \citep{Bond1984,fryer2001}. A numerous population of MBHs may have been the endproduct of the first episode of  pregalactic star formation; since they form in  density high peaks, relic MBHs with mass $\gta 150\,\msun$ would be predicted to cluster in the cores of more massive halos formed by subsequent mergers  \citep{MadauRees2001}.

Although this path to MBH formation seems very natural, large uncertainties exist on the final mass of PopIII stars.  If the first stars formed in multiples per halo the mass function would be less top-heavy \citep{Glover2008}. And, if single objects are formed, the characteristic final mass at the end of the accretion process can be much less than the initial clump mass due to feedback effects \citep{McKee2008}. 


\subsection{Gas-dynamical processes}
\label{ssec:1b}
Another family of models for MBH formation relies on the collapse of supermassive objects formed directly out of dense gas \citep[e.g.,][]{BrommLoeb2003, BVR2006,LN2006}. The physical conditions (density, gas content) in the inner regions of mainly gaseous proto-galaxies make these locii natural candidates, because the very first proto-galaxies were by definition metal-free, or at the very least very metal-poor. Enriched halos have a more efficient cooling, which in turn favors fragmentation and star formation over the efficient collection of gas conducive to MBH formation. It has been suggested that efficient gas collapse probably occurs only in massive halos with virial temperatures $T_{\rm vir} \gta 10^4$K under metal-free conditions where the formation of H$_2$ is inhibited \citep{BrommLoeb2003}, or for gas enriched below the critical metallicity threshold for fragmentation \citep{santoro}. For these systems the tenuous gas cools down by atomic hydrogen only until it reaches $T_{\rm gas}\sim 4000$ K.  At this point the cooling function of atomic hydrogen drops by a few orders of magnitude, and contraction proceeds nearly adiabatically. Finally, it has been recently suggested that highly turbulent systems are also likely to experience a limited amount of fragmentation, suggesting that efficient gas collapse could proceed also in metal-enriched galaxies at later cosmic epochs \citep{Begelman2009}. 

In such halos where fragmentation is suppressed, and cooling proceeds gradually,  the gaseous component can cool and contract until rotational support halts the collapse.   In the most common situations, rotational support can halt the collapse before densities required for MBH formation are reached. Additional mechanisms inducing transport of angular momentum are needed to further condense the gas until conditions fostering MBH formation are achieved. An appealing route to efficient angular momentum shedding is by global dynamical instabilities, such as the ``bars-within-bars" mechanism \citep{Shlosman1989, BVR2006}.  A bar can transport angular momentum outward on a dynamical timescale via gravitational and hydrodynamical torques, allowing the radius to shrink.  Provided that the gas is able to cool, this shrinkage leads to even greater instability, on shorter timescales, and the process cascades.  This mechanism is a very attractive candidate for collecting gas in the centers of halos, because it works on a dynamical time and can operate over many decades of radius. 

An alternative way of describing mass-inflow is via the Toomre stability parameter, $Q$. When $Q$ approaches unity a disc is subject to gravitational instabilities. If the destabilization of the system is not too violent, instabilities lead to mass infall instead of fragmentation and global star formation \citep{LN2006}.  This is the case if the inflow rate is below a critical threshold $\dot{M}_{ max}=2\alpha_{c}\frac{c^3_s}{G}$ the disk is able to sustain ($\alpha_c\,\sim\, 0.12$ describes the viscosity) and molecular and metal cooling are not important.  Such an unstable disc develops non-axisymmetric spiral structures, which effectively redistribute angular momentum, causing mass inflow.   This process stops when the amount of mass transported to the center is enough to make the disc marginally stable.

After gas has efficiently accumulated in the center, what happens next? The gas made available in the central compact region can then form a central massive object. Depending on how fast and efficiently the mass accumulation proceeds, the exact outcome would differ.  It is generally thought that VMSs supported by radiation pressure evolve as an $n=3$ polytrope \citep{hoyle1963,Baumgarte1999,saijo2002}. The fate of a marginally unstable, maximally rotating VMS has been investigated numerically in full general relativity by \cite{Shibata2002}. They found that the final object is a Kerr-like black hole (spin parameter $\approx$0.75) containing 90\% of the stellar mass. The fate of an isolated VMS is therefore the formation of a MBH. 

If the mass accumulation is fast, however, the outer layers of  VMSs are not thermally relaxed during much of the main sequence lifetime of the star \citep{Begelman2009b}. Such VMSs can have a complex structure with a convective (polytropic) core surrounded by a convectively stable envelope. After exhausting its hydrogen, the core of a VMS will contract and heat up until it suffers catastrophic neutrino losses and collapses. The initial black hole, with mass of a few $\msun$, formed as a result of core-collapse subsequently grows via accretion from the bloated envelopes that result  \citep[`quasistars', an initially low-mass black hole embedded within a massive, radiation-pressure-supported envelope; see also][]{BVR2006}. Over time, the black hole grows at the expense of the envelope, until finally the growing luminosity succeeds in unbinding the envelope and the seed MBH is unveiled. The key feature of this scenario is that while the black hole is embedded within the envelope, its growth is limited by the Eddington limit for the whole quasistar, rather than that appropriate for the black hole mass itself. Very rapid growth can then occur at early times, when the envelope mass greatly exceeds the black hole mass. 

The masses of the seeds predicted by different models of gas infall and VMS structure vary, but they are typically in the range $M_{BH} \sim 10^4-10^5\,M_\odot$ (Figure~\ref{fig:MF}). 

\subsection{Stellar-dynamical processes}
\label{ssec:1c}
Efficient gas collapse, leading to MBH seed formation, is mutually exclusive with star formation, as competition for the gas supply limits the mass available.  Formation of `normal' low--mass stars, however, opens up a new scenario for MBH seed formation, if  stellar-dynamical rather than gas-dynamical processes are at play.  This first episode of efficient star formation can  foster the formation of very compact nuclear star clusters \citep{schneider2006,clark2008,omukai2008} where star collisions can lead to the formation of a VMS, possibly leaving a MBH remnant with mass in the range $\sim\,10^2-10^4\,M_{\odot}$.
 \cite{Gaburov2009} investigate the process of runaway collisions directly with hydrodynamical simulations  and show that during  collisions large mass losses are likely, at metallicities larger than 10$^{-3}$ solar.  
 
%

The growth of a VMS should be much more efficient at low metallicity.  Low metallicity can modify the picture in different ways.  First,  at sub-solar (but non-zero) metallicity, stars with masses $\gta40M_\odot$ are thought to collapse directly into a black hole without exploding as supernovae.  Second, the mass loss due to winds is much more reduced in metal-poor stars, which greatly helps in increasing the mass of the final remnant.

\cite{Devecchi2009} investigate the formation of MBHs, remnants of VMS formed via stellar collisions in the very first stellar clusters at early cosmic times.  The main features of the model are as follows: we consider halos with virial temperatures $T_{vir}\simgt10^4$ K  after  the first episode of star formation, hence with a low, but non-zero, metallicity gas content.  This set of assumptions  ensures that (i) atomic hydrogen cooling can contribute to the gas  cooling process, (ii) a UV field has been created by the first stars,  and (iii) the gas inside the halo has been mildly polluted by the  first metals.  The second condition implies that at low density H$_2$  is dissociated and does not contribute to cooling. The third condition ensures that gas can fragment and form low-mass stars only if the gas density is above a certain threshold, $n_{crit,Z}$ \citep[which depends on the metallicity, see][]{santoro}: this causes only the highest density regions of a proto-galactic disc  to be prone to star formation. 

In Toomre-unstable proto-galactic discs, such as those described in the previous section, instabilities lead to mass infall instead of fragmentation into bound clumps and global star formation in the entire disk. The gas inflow increases the central density, and within a certain, compact, region  $n>n_{crit,Z}$. Here star formation ensues and a dense star cluster is formed. At metallicities $\sim 10^{-4}-10^{-5}$ solar, the typical star cluster masses are of order $10^5\msun$ and the typical half mass radii $\sim 1$ pc.  Most star cluster therefore go into core collapse in $\simlt$ 3 Myr, and runaway collisions of stars form a VMS \citep[see][]{Shapiro2004}, leading to a MBH remnant. 

As the  metallicity of the Universe increases, the critical density for  fragmentation decreases and stars start to form in the entire  protogalactic disk.  As gas is consumed by star formation, the inflow of gas is no longer efficient, more extended clusters form, and  the core collapse timescale increases until it exceeds the main sequence lifetime of massive stars. \cite{Devecchi2009}  find that typically a fraction $\sim 0.05$ of protogalaxies at $z\sim10-20$  form black hole seeds, with masses $\sim 1000-2000 \msun$, leading to  a mass density in seeds of a few $\simeq10^2\msun/{\rm   Mpc}^{-3}$.  Most of the assumptions in \cite{Devecchi2009} have been conservative, but still the population of seeds is comparable to the case of Population III star remnants discussed, for instance, in \cite{VHM}. The fraction of high-redshift galaxies seeded with a MBH is about a factor of 10 below the direct collapse case presented in \cite{VLN2008}, where a seed was assumed to form with a 100\% efficiency whenever a protogalaxy disk was Toomre unstable.

\section{Observational signatures of massive black hole seeds}

\begin{figure}[th]
\begin{center}
 \includegraphics[width=3.4in]{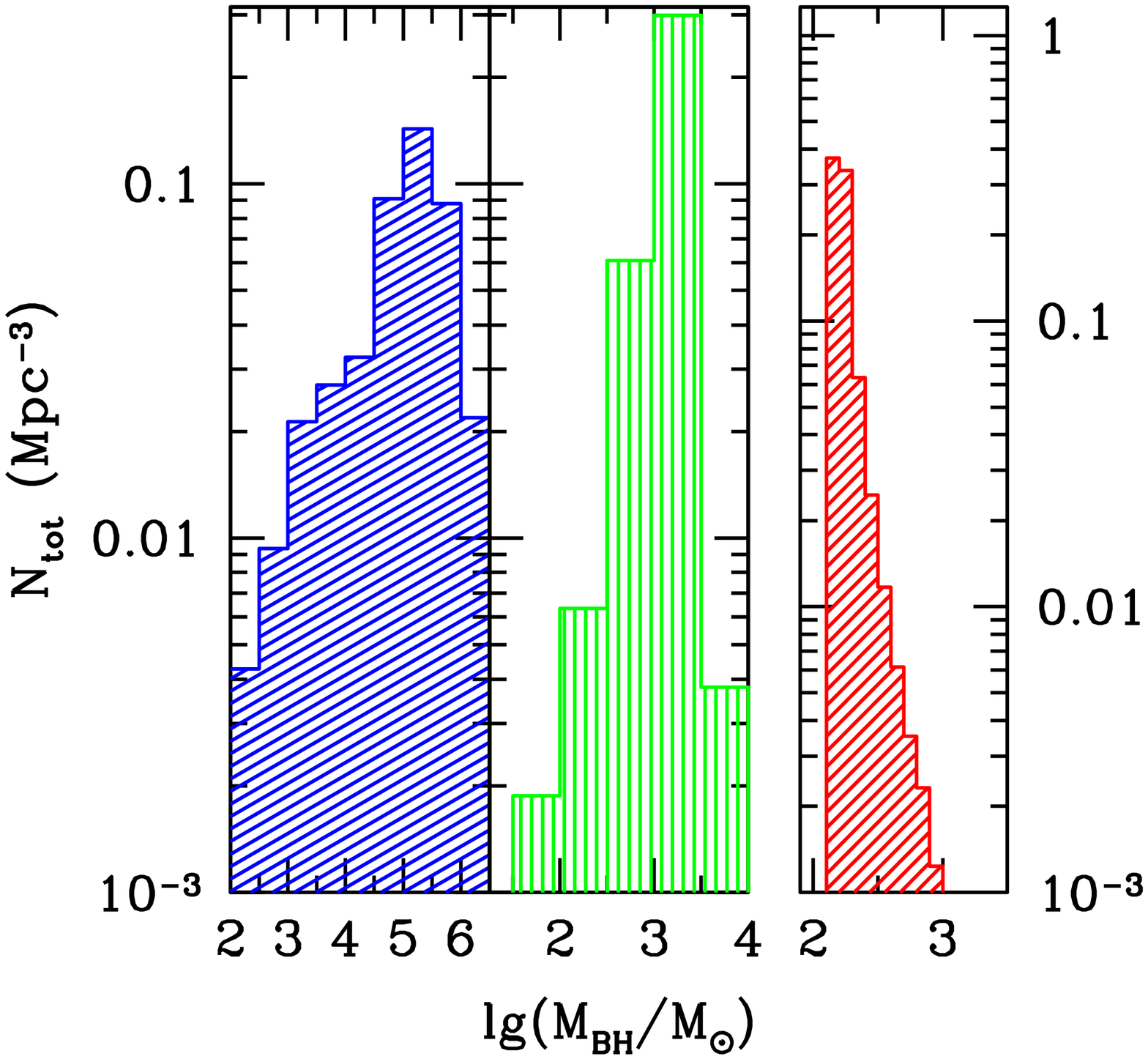} 
 \caption{Mass function of seed MBHs for three different formation scenarios: direct collapse \cite[left]{VLN2008}, runaway stellar mergers in high-redshift clusters \cite[center]{Devecchi2009}, and Population III remnants \cite[right]{MadauRees2001}. Note the different y-axis scale for the Population III case.}
   \label{fig:MF}
\end{center}
\end{figure}

What are the possible observational tests of MBH formation scenarios? Figure~\ref{fig:MF} shows three mass functions for three different MBH `seed' models: direct collapse \citep{VLN2008}, runaway stellar mergers in high-redshift clusters \citep{Devecchi2009}, and Population III remnants \citep{VHM}. These are the initial conditions that we would like to probe. In this section, I will focus on discriminants for two different, extreme, scenarios: `light seeds', derived from Population III remnants, and `heavy seeds', derived from direct gas collapse.

\subsection{Tracing MBHs at the earliest times}
Since during the quasar epoch ($z\approx 3-4$) MBHs increase their mass by a large factor \citep{YuTremaine2002}, signatures of the seed formation mechanisms are likely more evident at {\it earlier epochs}.  If accretion is efficient, it washes out any trace of the initial seed population. \cite{VG2009}  find that of the contribution to the quasar luminosity budget at $z>6$ is dominated by MBHs  with mass $<10^6\msun$. Such small, low-luminosity MBHs do not contribute to the bright end of the luminosity function of  quasars, and are therefore difficult to account from simple extrapolations of the luminosity  function of quasars. These small holes are not hosted in extremely massive galaxies residing  in the highest density peaks (5 to 6--$\sigma$ peaks), but are instead found  in more common,  ``normal'' systems, $\sim3-\sigma$, peaks.  Future generation of space--based telescopes, such as {\it JWST} and {\it IXO}, are likely to detect and constrain the evolution of the population of accreting massive black holes at early times ($z\simgt10$). 

\subsection{Gravitational waves}
Detection of gravitational waves from seeds merging at the redshift of formation \citep{GW3} is probably one of the best ways to discriminate among formation mechanisms. {\it LISA} in principle is sensible to gravitational waves from binary MBHs with masses in the range $10^3-10^6\; M_\odot$ basically at any redshift of interest.   A large fraction of coalescences will be directly observable by {\it LISA}, and on the basis of the detection rate, constraints can be put on the MBH formation process. 

\cite{GW3} and \cite{arun2009} analyze merger histories of MBHs along the  hierarchical build--up of cosmic structures \citep{VHM}.   \cite{GW3}  find that a decisive diagnostic is provided by the distribution  of the mass ratios in binary  coalescences. Models where seeds start large predict that most  of the detected events involve equal mass binaries. A fraction of observable coalescences, in fact, involve MBHs at $z>10$, when MBHs had no time to accrete  much mass yet. As most seeds form with similar mass, mergers at early times involve MBH binaries with mass ratio $\simeq 1$.  In scenarios based on Population III remnants, $z>10$ mergers involve MBHs with mass below the {\it LISA} threshold. The detectable events happen at later  times, when MBHs have already experienced a great deal of mass growth yielding a mass ratio distribution which is flat or features a broad  peak at mass ratios $\simeq 0.1-0.2$.   \cite{arun2009} determine the detectability of events by using  a code that includes both spin precession  and higher harmonics in the gravitational-wave signal, and carrying out Monte Carlo simulations to determine the number of events that can be detected  and accurately localized in these population models.   LISA will detect a significant fraction of all mergers: almost all mergers will be detected in heavy-seed scenarios, and nearly half of all mergers in small-seed scenarios.   For heavy-seed models most mergers detectable with high signal-noise-ratio (S/N$>$10) occur in the redshift range $3\simlt z \simlt 8$, with a peak around $z\sim 5$. In the case of light seeds, mergers are roughly uniform in $z$ over the range $4 \simlt z \simlt 10$.

\subsection{Massive black holes in low-mass galaxies: occupation fractions and $M_{\rm BH}-\sigma$}
The repercussions of different initial efficiencies for seed formation for the overall evolution of the MBH population stretch from high-redshift to the local Universe. The formation of seeds in a $\Lambda$CDM scenario follows the cosmological bias. As a consequence, the progenitors of massive galaxies (or clusters of galaxies) have a higher probability of hosting MBH seeds (cfr. \citealt{MadauRees2001}). In the case of low-bias systems, such as isolated low-mass galaxies, very few of the high-$z$ progenitors have the deep potential wells needed for gas retention and cooling, a prerequisite for MBH formation. 

\begin{figure}[h]
\begin{center}
 \includegraphics[width=3.4in]{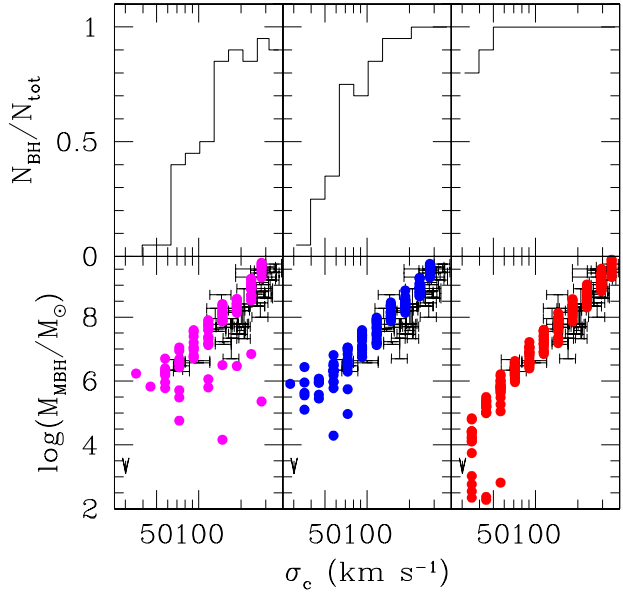} 
 \caption{The $M_{\rm bh}-$velocity dispersion ($\sigma_c$)
     relation at $z=0$. Every circle represents the central MBH in a
     halo of given $\sigma_c$.  Observational data are marked by their
     quoted errorbars, both in $\sigma_c$, and in $M_{\rm BH}$
     (Tremaine et al. 2002).  Left to right panels: $Q_{\rm c}=1.5$,
     $Q_{\rm c}=2$, Population III star seeds.  {\it
     Top panels:} fraction of galaxies at a given velocity dispersion
     which {\bf do} host a central MBH.}
   \label{fig3}
\end{center}
\end{figure}

The signature of the efficiency of the formation of MBH seeds will consequently be stronger in low-mass galaxies. Figure~\ref{fig3}
(bottom panel) shows a comparison between the observed $M_{\rm BH}-\sigma$ relation and the one predicted by different models (shown with circles), and in particular, from left to right, two models based on the \citet{LN2006} seed masses, and a third model based on lower-mass Population III star seeds. The upper panel of Figure~\ref{fig3} shows the fraction of galaxies that  do host a massive black hole for different velocity dispersion bins. This shows that the fraction of galaxies with a MBH increases with increasing halo masses at $z = 0$.  A larger fraction of low mass halos are devoid of central black holes for lower seed formation efficiencies. This is one of the key discriminants between different `seed' scenarios.

Simple arguments lead us to believe that MBHs might inhabit also the nuclei of dwarf galaxies, such as the satellites of the Milky Way and Andromeda, today (Van Wassenhove et al. 2009). Indeed, one of the best diagnostics of `seed' formation mechanisms would be to measure the masses of MBHs in dwarf galaxies. As MBHs grow from lower-mass seeds, it is natural to expect that a leftover population of progenitor MBHs should also exist in the present universe.  As discussed above, the progenitors of massive galaxies have a high probability that the central MBH is not ``pristine", that is, it has increased its mass by accretion, or it has experienced mergers and dynamical interactions. Any dependence of $\mbh$ on the initial seed mass is largely erased.  Dwarf galaxies undergo a quieter merger history, and as a result, at low masses the MBH occupation fraction and the distribution of MBH masses still retain some ``memory'' of the original seed mass distribution. The signatures of MBH formation mechanisms will consequently be stronger in  dwarf galaxies \citep{VLN2008}.  Van Wassenhove et al. (2009) find that for the most part MBHs hosted in Milky Way satellites retain the original `seed' mass, thus providing a clear indication of what the properties of the seeds were. MBHs generated as `massive seeds' have larger masses, that would favour their identification, their typical occupation fraction is lower, being always below 40\% and decreasing to less than a \% for `true' dwarf galaxy sizes.  Population III remnants have a higher occupation fraction, but their masses have not grown much since formation, making their detection harder. 

This can be understood by exploring the establishment and evolution of the empirical correlation between black hole mass ($M_{\rm BH}$) and velocity dispersion ($\sigma$) as a function of cosmic time in galaxies of different mass \citep{VN09}. The massive end of the $M_{\rm BH} - \sigma$ relation is established early, and lower mass MBHs migrate onto it as hierarchical merging proceeds.  The slope and scatter of the relation at the low-mass end instead retains memory of the seeding mechanism. We find that if MBH seeds are massive, $\sim 10^5\,M_{\odot}$, the low-mass end of the $\msigma$ flattens towards an asymptotic value, creating a characteristic `plume'. This `plume' consists of ungrown seeds, that merely continue to track the peak of the seed mass function down to late times.


\begin{thebibliography}{}

\bibitem[\protect\citeauthoryear{{Abel}, {Bryan} \& {Norman}}{{Abel}
  et~al.}{2000}]{abel2000}
{Abel} T.,  {Bryan} G.~L.,    {Norman} M.~L.,  2000, {ApJ}, 540, 39

\bibitem[\protect\citeauthoryear{{Arun}, {Babak}, {Berti}, {Cornish}, {Cutler},
  {Gair}, {Hughes}, {Iyer}, {Lang}, {Mandel}, {Porter}, {Sathyaprakash},
  {Sinha}, {Sintes}, {Trias}, {Van Den Broeck} \& {Volonteri}}{{Arun}
  et~al.}{2009}]{arun2009}
{Arun} K.~G.,  {Babak} S.,  {Berti} E.,  {Cornish} N.,  {Cutler} C.,  {Gair}
  J.,  {Hughes} S.~A.,  {Iyer} B.~R.,  {Lang} R.~N.,  {Mandel} I.,  {Porter}
  E.~K.,  {Sathyaprakash} B.~S.,  {Sinha} S.,  {Sintes} A.~M.,  {Trias} M.,
  {Van Den Broeck} C.,    {Volonteri} M.,  2009, Classical and Quantum Gravity,
  26, 094027

\bibitem[\protect\citeauthoryear{{Barth}, {Ho}, {Rutledge} \&
  {Sargent}}{{Barth} et~al.}{2004}]{barthetal2004}
{Barth} A.~J.,  {Ho} L.~C.,  {Rutledge} R.~E.,    {Sargent} W.~L.~W.,  2004,
  {ApJ}, 607, 90

\bibitem[\protect\citeauthoryear{{Barth}, {Martini}, {Nelson} \& {Ho}}{{Barth}
  et~al.}{2003}]{Barthetal2003}
{Barth} A.~J.,  {Martini} P.,  {Nelson} C.~H.,    {Ho} L.~C.,  2003, {ApJL},
  594, L95

\bibitem[\protect\citeauthoryear{{Baumgarte} \& {Shapiro}}{{Baumgarte} \&
  {Shapiro}}{1999}]{Baumgarte1999}
{Baumgarte} T.~W.,  {Shapiro} S.~L.,  1999, \apj, 526, 941

\bibitem[\protect\citeauthoryear{{Begelman}}{{Begelman}}{2009}]{Begelman2009b}
{Begelman} M.~C.,  2009, ArXiv e-prints

\bibitem[\protect\citeauthoryear{{Begelman} \& {Shlosman}}{{Begelman} \&
  {Shlosman}}{2009}]{Begelman2009}
{Begelman} M.~C.,  {Shlosman} I.,  2009, \apjl, 702, L5

\bibitem[\protect\citeauthoryear{{Begelman}, {Volonteri} \& {Rees}}{{Begelman}
  et~al.}{2006}]{BVR2006}
{Begelman} M.~C.,  {Volonteri} M.,    {Rees} M.~J.,  2006, {MNRAS}, 370, 289

\bibitem[\protect\citeauthoryear{{Bond}, {Arnett} \& {Carr}}{{Bond}
  et~al.}{1984}]{Bond1984}
{Bond} J.~R.,  {Arnett} W.~D.,    {Carr} B.~J.,  1984, \apj, 280, 825

\bibitem[\protect\citeauthoryear{{Bromm}, {Coppi} \& {Larson}}{{Bromm}
  et~al.}{1999}]{bromm1999}
{Bromm} V.,  {Coppi} P.~S.,    {Larson} R.~B.,  1999, {ApJL}, 527, L5

\bibitem[\protect\citeauthoryear{{Bromm}, {Coppi} \& {Larson}}{{Bromm}
  et~al.}{2002}]{bromm2002}
{Bromm} V.,  {Coppi} P.~S.,    {Larson} R.~B.,  2002, {ApJ}, 564, 23

\bibitem[\protect\citeauthoryear{{Bromm} \& {Loeb}}{{Bromm} \&
  {Loeb}}{2003}]{BrommLoeb2003}
{Bromm} V.,  {Loeb} A.,  2003, {ApJ}, 596, 34

\bibitem[\protect\citeauthoryear{{Clark}, {Glover} \& {Klessen}}{{Clark}
  et~al.}{2008a}]{Glover2008}
{Clark} P.~C.,  {Glover} S.~C.~O.,    {Klessen} R.~S.,  2008a, \apj, 672, 757

\bibitem[\protect\citeauthoryear{{Clark}, {Glover} \& {Klessen}}{{Clark}
  et~al.}{2008b}]{clark2008}
{Clark} P.~C.,  {Glover} S.~C.~O.,    {Klessen} R.~S.,  2008b, ApJ, 672, 757

\bibitem[\protect\citeauthoryear{{Devecchi} \& {Volonteri}}{{Devecchi} \&
  {Volonteri}}{2009}]{Devecchi2009}
{Devecchi} B.,  {Volonteri} M.,  2009, \apj, 694, 302

\bibitem[\protect\citeauthoryear{{Ferrarese} \& {Ford}}{{Ferrarese} \&
  {Ford}}{2005}]{ferrareseford}
{Ferrarese} L.,  {Ford} H.,  2005, Space Science Reviews, 116, 523

\bibitem[\protect\citeauthoryear{{Fryer}, {Woosley} \& {Heger}}{{Fryer}
  et~al.}{2001}]{fryer2001}
{Fryer} C.~L.,  {Woosley} S.~E.,    {Heger} A.,  2001, {ApJ}, 550, 372

\bibitem[\protect\citeauthoryear{{Gaburov}, {Lombardi} \& {Portegies
  Zwart}}{{Gaburov} et~al.}{2009}]{Gaburov2009}
{Gaburov} E.,  {Lombardi} J.,    {Portegies Zwart} S.,  2009, ArXiv e-prints

\bibitem[\protect\citeauthoryear{{Glover}, {Clark}, {Greif}, {Johnson},
  {Bromm}, {Klessen} \& {Stacy}}{{Glover} et~al.}{2008}]{Glover2008a}
{Glover} S.~C.~O.,  {Clark} P.~C.,  {Greif} T.~H.,  {Johnson} J.~L.,  {Bromm}
  V.,  {Klessen} R.~S.,    {Stacy} A.,  2008, in {Hunt} L.~K.,  {Madden} S.,
  {Schneider} R.,  eds, IAU Symposium Vol.~255 of IAU Symposium, {Open
  questions in the study of population III star formation}.
pp 3--17

\bibitem[\protect\citeauthoryear{{G{\"u}ltekin}, {Richstone}, {Gebhardt},
  {Lauer}, {Tremaine}, {Aller}, {Bender}, {Dressler}, {Faber}, {Filippenko},
  {Green}, {Ho}, {Kormendy}, {Magorrian}, {Pinkney} \& {Siopis}}{{G{\"u}ltekin}
  et~al.}{2009}]{Gultekin2009}
{G{\"u}ltekin} K.,  {Richstone} D.~O.,  {Gebhardt} K.,  {Lauer} T.~R.,
  {Tremaine} S.,  {Aller} M.~C.,  {Bender} R.,  {Dressler} A.,  {Faber} S.~M.,
  {Filippenko} A.~V.,  {Green} R.,  {Ho} L.~C.,  {Kormendy} J.,  {Magorrian}
  J.,  {Pinkney} J.,    {Siopis} C.,  2009, \apj, 698, 198

\bibitem[\protect\citeauthoryear{{Hoyle} \& {Fowler}}{{Hoyle} \&
  {Fowler}}{1963}]{hoyle1963}
{Hoyle} F.,  {Fowler} W.~A.,  1963, MNRAS, 125, 169

\bibitem[\protect\citeauthoryear{{Kudritzki} \& {Puls}}{{Kudritzki} \&
  {Puls}}{2000}]{Kudritzki2000}
{Kudritzki} R.-P.,  {Puls} J.,  2000, \araa, 38, 613

\bibitem[\protect\citeauthoryear{{Lodato} \& {Natarajan}}{{Lodato} \&
  {Natarajan}}{2006}]{LN2006}
{Lodato} G.,  {Natarajan} P.,  2006, {MNRAS}, 371, 1813

\bibitem[\protect\citeauthoryear{{Madau} \& {Rees}}{{Madau} \&
  {Rees}}{2001}]{MadauRees2001}
{Madau} P.,  {Rees} M.~J.,  2001, {ApJL}, 551, L27

\bibitem[\protect\citeauthoryear{{McKee} \& {Tan}}{{McKee} \&
  {Tan}}{2008}]{McKee2008}
{McKee} C.~F.,  {Tan} J.~C.,  2008, \apj, 681, 771

\bibitem[\protect\citeauthoryear{{Omukai}, {Schneider} \& {Haiman}}{{Omukai}
  et~al.}{2008}]{omukai2008}
{Omukai} K.,  {Schneider} R.,    {Haiman} Z.,  2008, \apj, 686, 801

\bibitem[\protect\citeauthoryear{{Palla}, {Zinnecker}, {Maeder} \&
  {Meynet}}{{Palla} et~al.}{2002}]{palla2002}
{Palla} F.,  {Zinnecker} H.,  {Maeder} A.,    {Meynet} G.,  eds, 2002, {Physics
  of star formation in galaxies}

\bibitem[\protect\citeauthoryear{{Peterson}, {Bentz}, {Desroches},
  {Filippenko}, {Ho}, {Kaspi}, {Laor}, {Maoz}, {Moran}, {Pogge} \&
  {Quillen}}{{Peterson} et~al.}{2005}]{Peterson2005}
{Peterson} B.~M.,  {Bentz} M.~C.,  {Desroches} L.,  {Filippenko} A.~V.,  {Ho}
  L.~C.,  {Kaspi} S.,  {Laor} A.,  {Maoz} D.,  {Moran} E.~C.,  {Pogge} R.~W.,
   {Quillen} A.~C.,  2005, \apj, 632, 799

\bibitem[\protect\citeauthoryear{{Rees}}{{Rees}}{1978}]{Rees1978}
{Rees} M.~J.,  1978, in {Berkhuijsen} E.~M.,  {Wielebinski} R.,  eds, Structure
  and Properties of Nearby Galaxies Vol.~77 of IAU Symposium, {Emission from
  the nuclei of nearby galaxies - Evidence for massive black holes}.
pp 237--242

\bibitem[\protect\citeauthoryear{{Saijo}, {Baumgarte}, {Shapiro} \&
  {Shibata}}{{Saijo} et~al.}{2002}]{saijo2002}
{Saijo} M.,  {Baumgarte} T.~W.,  {Shapiro} S.~L.,    {Shibata} M.,  2002, ApJ,
  569, 349

\bibitem[\protect\citeauthoryear{{Santoro} \& {Shull}}{{Santoro} \&
  {Shull}}{2006}]{santoro}
{Santoro} F.,  {Shull} J.~M.,  2006, ApJ, 643, 26

\bibitem[\protect\citeauthoryear{{Schneider}, {Omukai}, {Inoue} \&
  {Ferrara}}{{Schneider} et~al.}{2006}]{schneider2006}
{Schneider} R.,  {Omukai} K.,  {Inoue} A.~K.,    {Ferrara} A.,  2006, MNRAS,
  369, 1437

\bibitem[\protect\citeauthoryear{{Sesana}, {Volonteri} \& {Haardt}}{{Sesana}
  et~al.}{2007}]{GW3}
{Sesana} A.,  {Volonteri} M.,    {Haardt} F.,  2007, MNRAS, 377, 1711

\bibitem[\protect\citeauthoryear{{Shapiro}}{{Shapiro}}{2004}]{Shapiro2004}
{Shapiro} S.~L.,  2004, in {Ho} L.~C.,  ed., Coevolution of Black Holes and
  Galaxies {Formation of Supermassive Black Holes: Simulations in General
  Relativity}.
pp 103--+

\bibitem[\protect\citeauthoryear{{Shibata} \& {Shapiro}}{{Shibata} \&
  {Shapiro}}{2002}]{Shibata2002}
{Shibata} M.,  {Shapiro} S.~L.,  2002, \apjl, 572, L39

\bibitem[\protect\citeauthoryear{{Shlosman}, {Frank} \& {Begelman}}{{Shlosman}
  et~al.}{1989}]{Shlosman1989}
{Shlosman} I.,  {Frank} J.,    {Begelman} M.~C.,  1989, Nature, 338, 45

\bibitem[\protect\citeauthoryear{{Volonteri} \& {Gnedin}}{{Volonteri} \&
  {Gnedin}}{2009}]{VG2009}
{Volonteri} M.,  {Gnedin} N.~Y.,  2009, \apj, 703, 2113

\bibitem[\protect\citeauthoryear{{Volonteri}, {Haardt} \& {Madau}}{{Volonteri}
  et~al.}{2003}]{VHM}
{Volonteri} M.,  {Haardt} F.,    {Madau} P.,  2003, {ApJ}, 582, 559

\bibitem[\protect\citeauthoryear{{Volonteri}, {Lodato} \&
  {Natarajan}}{{Volonteri} et~al.}{2008}]{VLN2008}
{Volonteri} M.,  {Lodato} G.,    {Natarajan} P.,  2008, MNRAS, 383, 1079

\bibitem[\protect\citeauthoryear{{Volonteri} \& {Natarajan}}{{Volonteri} \&
  {Natarajan}}{2009}]{VN09}
{Volonteri} M.,  {Natarajan} P.,  2009, \mnras, pp 1476--+

\bibitem[\protect\citeauthoryear{{Yoshida}, {Omukai}, {Hernquist} \&
  {Abel}}{{Yoshida} et~al.}{2006}]{Yoshida2006}
{Yoshida} N.,  {Omukai} K.,  {Hernquist} L.,    {Abel} T.,  2006, \apj, 652, 6

\bibitem[\protect\citeauthoryear{{Yu} \& {Tremaine}}{{Yu} \&
  {Tremaine}}{2002}]{YuTremaine2002}
{Yu} Q.,  {Tremaine} S.,  2002, {MNRAS}, 335, 965

\end{thebibliography}
\end{document}